\documentclass[pra, twocolumn,amsmath,amssymb,superscriptaddress]{revtex4-1}

\usepackage{amsfonts}
\usepackage{physics}
\usepackage{graphicx}
\usepackage{natbib}
\usepackage{tikz}
\usepackage{soul}
\usepackage{dutchcal}

\newcommand{\be}{\begin{equation}}
\newcommand{\ee}{\end{equation}}

\newcommand{\bea}{\begin{eqnarray}}
\newcommand{\eea}{\end{eqnarray}}

\newcommand{\sn}[1]{_{\text{\tiny $#1$}}}

\newcommand{\pD}[2]{\frac{\partial #1}{\partial #2}}

\newcommand{\fourmatrix}[4]{\left(\begin{array}{rr} #1 & #2\\ #3 & #4\end{array}\right)}
\newcommand{\twovector}[2]{\left(\begin{array}{c} #1 \\ #2 \end{array}\right)}

\begin{document}

\title{Trapped-particle evolution driven by residual gas collisions}
\author{Avinash Deshmukh}
\email{aud2@phas.ubc.ca}
\affiliation{Department of Physics \& Astronomy, University of British Columbia, 6224 Agricultural Road, Vancouver, B.C., V6T 1Z1, Canada} 

\author{Riley A. Stewart}
\affiliation{Department of Physics \& Astronomy, University of British Columbia,  6224 Agricultural Road, Vancouver, B.C., V6T 1Z1, Canada}
\author{Pinrui Shen}
\affiliation{Department of Physics \& Astronomy, University of British Columbia, 6224 Agricultural Road, Vancouver, B.C., V6T 1Z1, Canada} 
\author{James L. Booth} 
\affiliation{Physics Department, British Columbia Institute of Technology, 3700 Willingdon Avenue, Burnaby, B.C. V5G 3H2, Canada}

\author{Kirk W. Madison}
\email{madison@phas.ubc.ca}
\affiliation{Department of Physics \& Astronomy, University of British Columbia, 6224 Agricultural Road, Vancouver, B.C., V6T 1Z1, Canada}

\date{\today}

\begin{abstract}
\noindent{We present a comprehensive mathematical model and experimental measurements for the evolution of a trapped particle ensemble driven by collisions with a room-temperature background vapor. The model accommodates any trap geometry, confining potential, initial trapped distribution, and other experimental details; it only depends on the probability distribution function $P_t(E)$ for the collision-induced energy transfer to the trapped ensemble. We describe how to find $P_t(E)$ using quantum scattering calculations and how it can be approximated using quantum diffractive universality \cite{Booth2019}. We then compare our model to experimental measurements of a $^{87}$Rb ensemble energy evolution exposed to a room temperature background gas of Ar by means of a single parameter fit for the total collision rate $\Gamma$.  We extracted a collision rate of $\Gamma = 0.649(2)\ \text{s}^{-1}$. We further refine our analysis by using monotonic Gaussian process regression to smooth the experimental data, which extracts a collision rate of $\Gamma = 0.646(1)\ \text{s}^{-1}$. This is compared to a value of 0.67(1) $\text{s}^{-1}$ found by the commonly used method of zero-trap depth extrapolation, a 3.5\% correction that is a result of our model fully taking ensemble loss and heating into account. Finally, we report a five-fold increase in the precision of our collision rate extraction from the experimental data and a ten-fold increase in the precision of our collision rate extraction from the smoothed experimental data.}
\end{abstract}

\pacs{34.50.-s, 34.50.Cx, 34.80.Bm, 34.00.00, 30.00.00, 67.85.-d, 37.10.Gh}

\maketitle

\section{Introduction}

\noindent{In a vacuum, single particles or ensembles of single particles are constantly perturbed by collisions with atoms and molecules in the residual background vapor.  Energetic elastic and inelastic collisions can result in neutral particle loss from traps \cite{Migdall1985,Bjorkholm1988,Hemmerling_2014,PhysRevLett.112.023002} and ion replacement in ion traps \cite{10.1063/1.5104346}.  Collisions that do not impart enough energy to free a trapped particle also constitute an important decoherence mechanism.   These collisions can induce heating in neutral ensembles \cite{PhysRevA.61.033606,PhysRevA.62.063614,cornell-1999,PhysRevA.60.R29}, melting of ion Coulomb crystals \cite{PhysRevA.105.052426,Pagano_2019,PRXQuantum.4.020317}, reordering of ion chains \cite{10.1063/5.0029236,9606562}, decoherence in trapped-atom \cite{PhysRevLett.110.180802} and trapped-ion optical atomic clocks \cite{PhysRevA.100.033419}, and localization-induced decoherence of single levitated nanoparticles \cite{magrini_real-time_2021,Aspelmeyer-2023}. Understanding the nature of these collisions is important for both producing cold particle ensembles and using them in quantum sensing and quantum information applications.}\\

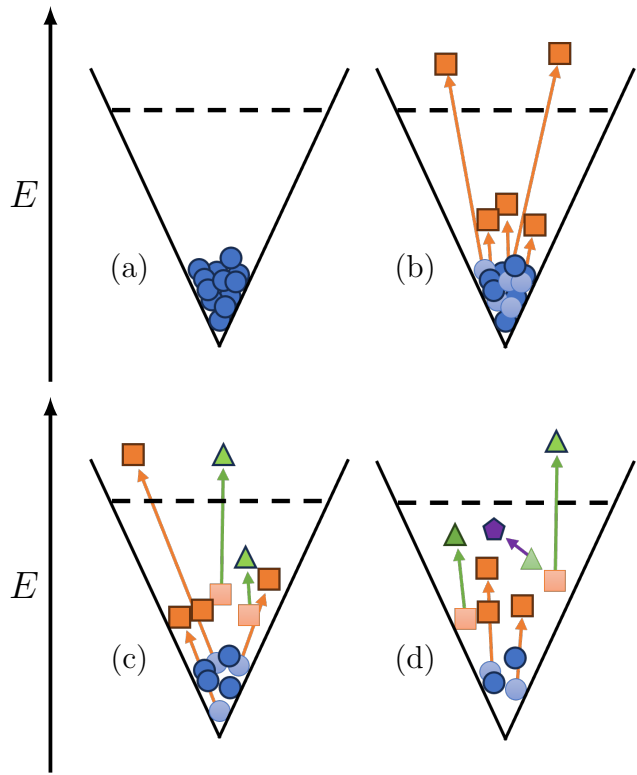
\begin{figure}
	\centering
 \begin{tikzpicture}
        \node at (-2.6,0) {\Large $E$};
        \draw[-latex, line width = 0.5mm] (-2.25,-2.5)--(-2.25,2.5) ;
		\node at (0,0){\includegraphics[width= 0.4\linewidth]{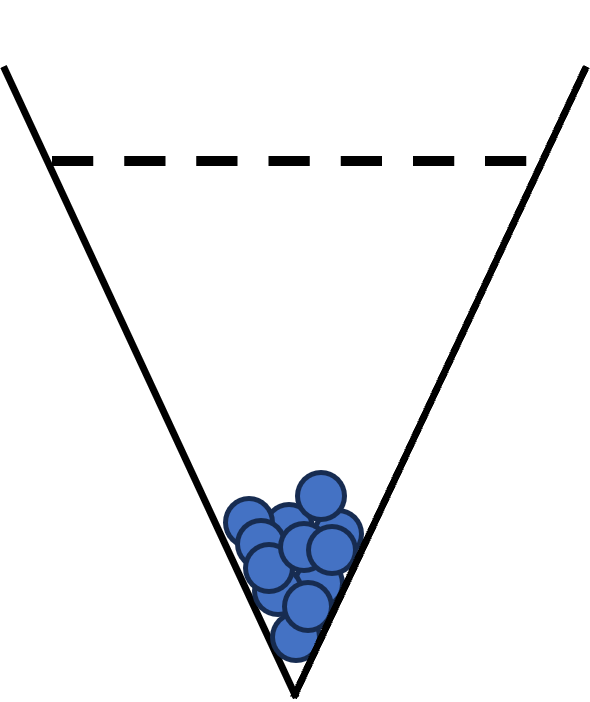}};
        \node at (-1.2,-1){\large (a)};
		\node at (3.8,0) {\includegraphics[width = 0.4\linewidth]{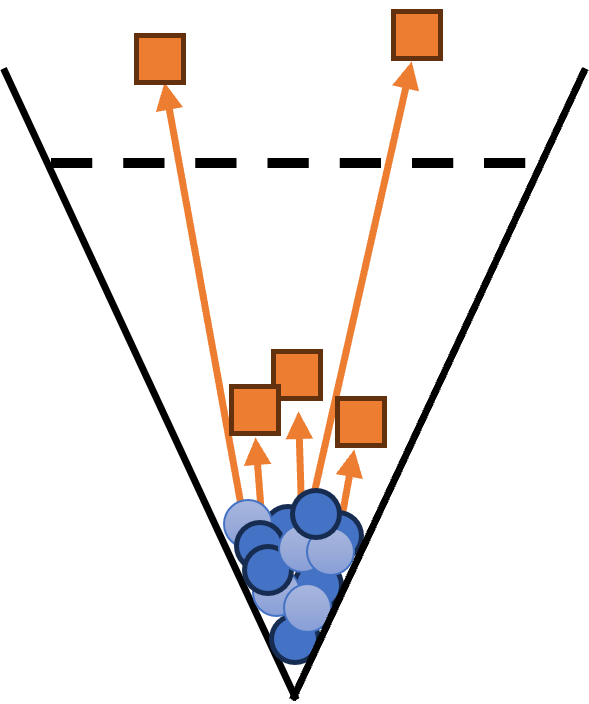}};
        \node at (2.6,-1){\large (b)};
          \node at (-2.6,-5.2) {\Large $E$};
        \draw[-latex, line width = 0.5mm] (-2.25,-7.7)--(-2.25,-2.7) ;
       \node at (-1.2,-6.2){\large (c)};
		\node at (0,-5.2){\includegraphics[width= 0.4\linewidth]{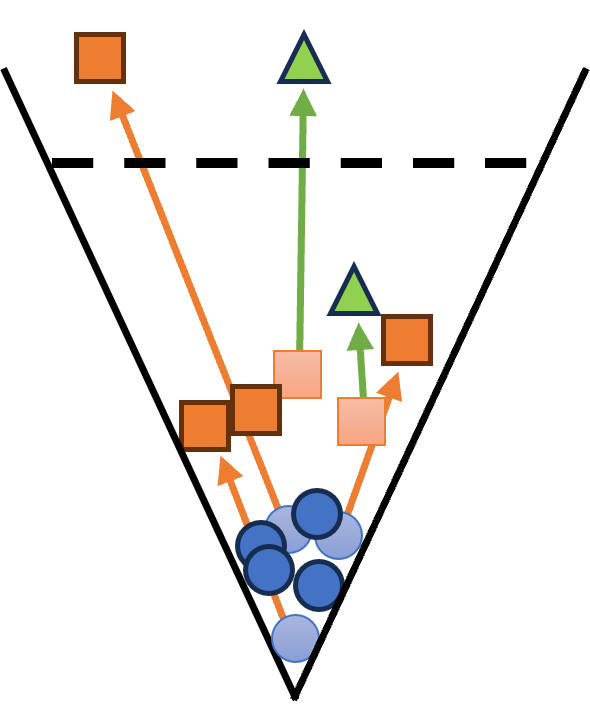}};
          \node at (2.6,-6.2){\large (d)};
		\node at (3.8,-5.2) {\includegraphics[width = 0.4\linewidth]{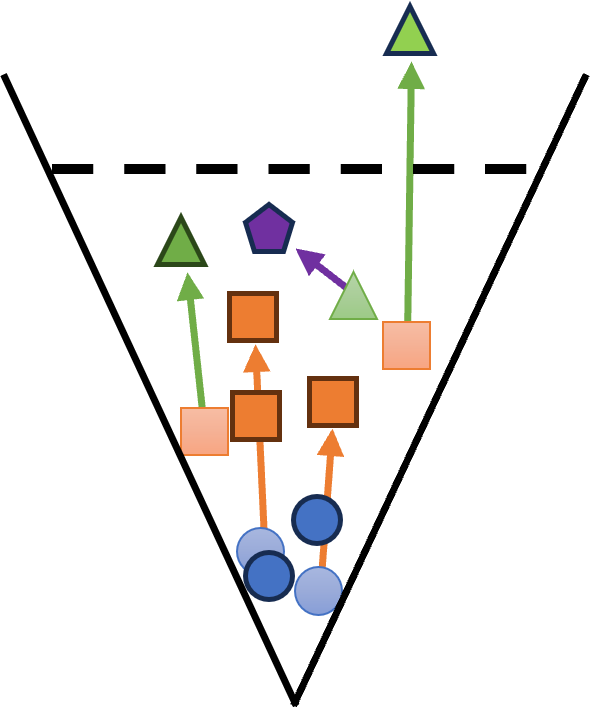}};
\end{tikzpicture}

        \caption{A schematic showing how background collisions cause the trapped energy ensemble to evolve. The vertical axis denotes energy and the horizontal axis is spatial position in a linear trap potential; however, the form of the confining potential is irrelevant to the analysis presented. The dashed horizontal line demarcates the maximum trap depth $E_m$. (a) Upon loading in the trap, all of the particles (solid blue circles) have a low energy and have not undergone any collisions with the background. (b) At short times, some of the particles (lightly shaded circles) have undergone a single collision, which causes them to go to a new state (solid orange squares). Particles with a new energy larger than $E_m$ are lost. Those with less remain in the trap, but with a higher energy than before -- the ensemble has heated. (c) At longer times, particles continue to undergo collisions from a zero to one collision state and also from a one to two collision state (solid green triangles). The one collision particles have a higher energy than they had at the initial time, leading to a higher loss rate. (d) At long times, the energy distribution bears very little similarity to the initial distribution. More particles are in collided states than in initial states, and there are some particles that have undergone three collisions (solid purple pentagon). These particles can have an energy very close to $E_m$, and thus have a very small probability of being retained after a fourth collision. This allows the truncation of our analysis after a few collision orders.}
	\label{Trap_Schematic}
 
\end{figure}

While understanding the qualitative role of neutral atom loss and ensemble heating due to collisions with the residual background vapor was important for achieving Bose-Einstein condensation in alkali metal gases \cite{PhysRevLett.70.414,cornell-1999}, a rigorous and quantitative understanding of the two effects is essential for the emerging field of quantum vacuum metrology.  Here, precise measurements of a sensor particle's collision rate $\Gamma= n \langle \sigma(v) v \rangle$ can be used with knowledge of the thermally-averaged, velocity-weighted collision cross section $\langle\sigma(v)v\rangle$ to determine the background particle density $n$ and subsequent background pressure \cite{Booth2011,Madison2012}.  Previous measurements of sensor atom collision rates were conducted by observing the sensor-atom loss rate from shallow magnetic \cite{PhysRevA.80.022712,PhysRevA.84.022708,Julia2017,Julia2018,Eckel_2018,Booth2019,Shen_2020,Shen_2021,PhysRevA.106.052812,doi:10.1116/5.0095011,Shen_2023,10.1116/5.0147686} and optical dipole traps \cite{Makhalow2016,Makhalov_2017}.  As illustrated in Figure \ref{Trap_Schematic}, the challenge with this approach lies in the fact that even for exceedingly small trap depths, not every collision results in loss \cite{PhysRevA.80.022712,PhysRevA.84.022708}. This complicates extracting the particle's collision rate from the observed trap loss rate as the instantaneous loss rate varies in time due to its energy dependence.  Indeed, both the instantaneous loss rate and heating rate depend on the maximum trap depth (defined for an atom with zero energy) and on the energy distribution of the sensor atoms in the trap. Collisions that do not eject sensor atoms change (heat) the energy distribution of the sensor ensemble, causing the effective trap depth and corresponding loss rate to vary with time.  This means that a simple exponential loss model does not adequately capture the behavior of trap loss.\\

As we discuss in Appendix A, extrapolation of this loss rate to zero trap depth to obtain the total collision rate relies on data with very low signal-to-noise ratio (SNR), because low trap depth measurements requires culling high energy atoms. These measurements subsequently have a large uncertainty on the total collision rate. This problem cannot be corrected by using higher trap depth measurements, because the variation in the loss rate sharply decreases at small trap depths. This flattening of the loss rate curve cannot be observed at larger trap depths, which makes extrapolation from large trap depth inaccurate. There is the additional complication that this method involves a sequential multidimensional fit; the first being a fit of a trap population curve as a function of time, and the second being an extrapolation of the resulting loss rates as a function of energy. This introduces additional fit uncertainty, and, because the fits are done sequentially, neglects any kind of covariance in the two fit dimensions. Our recent work has accounted for the role of the sensor-atom energy distribution and has added heating corrections to loss rate measurements \cite{PhysRevA.106.052812,Shen_2021,Shen_2023}; however, a more complete description requires an analysis of how the trapped particle energy distribution evolves in time due to these collisions.\\

In this work, we present a comprehensive theoretical treatment of trapped particle loss and heating due to collisions with a room-temperature background vapor \cite{Deshmukh2023}.  We derive an analytic expression for the trapped particle energy distribution evolution in time based solely on a measured ensemble energy distribution (as the initial condition), the total collision rate, and the probability distribution function $P_t(E)$ for the energy transferred in a collision. Our treatment differs from prior work on the trap depth dependence of ensemble heating rates \cite{PhysRevA.61.033606,PhysRevA.62.063614,PhysRevA.106.052812,Shen_2021,Shen_2023} in that it is valid for all time, for any initial energy distribution, and fully corrects analysis for heating of the trap. Moreover, it does not rely on a particular form for the transferred energy distribution, and is general with respect to trap-background interactions. Finally, our method can extract a collision rate using a single dimensional fit that can be made more precise by smoothing high SNR data instead of extrapolating low SNR data. We present a direct comparison of the theory with experimental measurements of a magnetically trapped rubidium-87 ensemble exposed to a background vapor of argon.  By analyzing the energy evolution of the particles in the trap, we determine the total collision rate with a precision of better than 0.2\%. This represents a significant improvement (more than a factor of 5) in the precision for the total collision rate determination. Our theory was recently applied in Ref.~\cite{Eckel2023} to correct the total collision rate reported in a comparison between two cold atom vacuum standards and a classical vacuum metrology apparatus \cite{10.1116/5.0147686}.

\section{The Energy Distribution Evolution Model}

\noindent{We consider an ensemble of non-interacting particles with mass $m$ subject to a spatially dependent trap potential $U$. A particle with initial energy $E_i$ and momentum $\textbf{p}_0$, subject to $k$ collisions which transfer  momenta $\textbf{p}_{1},...,\textbf{p}_k$ to the particle respectively, will have a final energy $E_f$ given by}

	\begin{equation}
 \label{K-Collision Energies}
		E_f = E_i + \sum_{k=1}^{n} \left(E_k + \dfrac{\textbf{p}_k}{m}\cdot \sum_{i=0}^{k-1} \textbf{p}_i\right),
	\end{equation}

\noindent{where $E_k = \textbf{p}_k^2/2m$. Details are shown in Appendix B. Since the probability of a collision event is independent of the number of previous collisions undergone by the particle, we model the collisions as Poissonian events that occur with a constant collision rate $\Gamma$. The fraction $\rho_k$ of the ensemble that will have undergone $k$ collisions in a time $t$ is then}
	\begin{equation}\label{PoissionWeights}
		\rho_k(\Gamma t) = \dfrac{(\Gamma t)^k e^{-\Gamma t}}{k!}.
	\end{equation}
	
	\noindent{The energy distribution of the ensemble $f(E,t)$ at time $t$ will be the sum of the energy distributions $f_k(E)$ corresponding to the subset of particles that underwent exactly $k$ collisions weighted by the Poissonian probability that $k$ collision events occurred. We can write this as \cite{Deshmukh2023}}
	
 	\begin{equation}\label{DistributionPoissonianSum}
		f(E,t) = \sum_{k=0}^{\infty} \dfrac{(\Gamma t)^k e^{-\Gamma t}}{k!} f_k(E).
	\end{equation}
	
	{The distribution $f_k(E)$ can be computed using the initial energy distribution $f_0(E)$ and the probability $P_t(E)$ that a collision transfers an energy $E$.  Since the incident velocities of the background particles are isotropic, the momentum imparted by the collisions is also isotropic. Then, if we average over the ensemble of particles at a certain energy, the contribution of the cross-term $\textbf{p}\cdot\textbf{p}_t/m$ appearing in Eqn.~\ref{K-Collision Energies} will vanish. Having appropriate cross-terms such that $E_f < E_i$ is the only way we can have a collision that cools the trapped particle, so we do not need to consider such collisions here. Explicitly, this means that $P_t(E)$ vanishes for $E < 0$. We can therefore perform $k$ convolutions of $f_0(E)$ with $P_t(E)$ for all positive energies to find the distribution $f_k(E)$. We can write this as} 
	
	\begin{align}\label{fkEquation}
		f_k(E) = \int_0^{\infty} & dE_1\ P_t(E-E_1) \\ &\times \prod_{j=2}^{k} \left(\int_0^{\infty} dE_j P_t(E_{j-1} - E_j) \right) f_0(E_k)\nonumber,
	\end{align}
	
	\noindent{where $P_t(E<0)=0$ and the product denotes iterated integrals. Substituting this into our expression for the full distribution, we have}
	
	\begin{align}
		f(E,t) = \sum_{k=0}^{\infty} & \dfrac{(\Gamma t)^k e^{-\Gamma t}}{k!} \int_0^{\infty} dE_1\ P_t(E-E_1) \label{PDF_Model_Equation}\\ 
 & \times \prod_{j=2}^{k} \left(\int_0^{\infty} dE_j P_t(E_{j-1} - E_j) \right) f_0(E_k)\nonumber.
	\end{align}

 \noindent{Sample plots showing how the total energy distribution and the populations of the $f_k(E)$ distributions change in time are shown in Fig. \ref{Sub-Ensemble_Plots}.}\\

{In traps of finite depth, particles above some energy $U$ are not trapped, and all collisions that increase the energy above this trap depth result in loss. In this case, we only convolve the distributions up to the energy $U$, so the previous expression becomes}

\begin{align}
		f(E,t) = & \Theta(U-E)\sum_{k=0}^{\infty} \dfrac{(\Gamma t)^k e^{-\Gamma t}}{k!} \int_0^{U} dE_1\ P_t(E-E_1)\label{ModelExpression}\nonumber\\ 
 & \times  \prod_{j=2}^{k} \left(\int_0^{U} dE_j P_t(E_{j-1} - E_j) \right) f_0(E_k).
	\end{align}

 \indent{Because $P_t(E)$ does not necessarily vanish for $E > U$, the convolution up to $U$ does not result in a normalized distribution. This reflects loss from the trap; particles that are transferred an energy greater than the trap depth will be lost from the trap and will not appear in experimentally measured energy distributions. We therefore see that Eqn. \ref{ModelExpression} provides the trapped particle energy distribution accounting for both heating and loss.}\\

 \indent{Our analysis assumes that the particles in the gas do not interact, which means that the distribution of the trapped ensemble does not change due to intratrap collisions (collisions between two trapped particles). As such, we can neglect the evolution of the distribution due to thermalization. This assumption holds when the intratrap collision rate is much lower than the background collision rate. This is the case in vacuum sensor measurements, where the trapped ensemble is sufficiently diffuse and cold. Further discussion can be found in Appendix C. In the case where the intratrap collision time scale is close to or less than the background collision time scale, the theory must be modified. It is well-known that in the absence of background collisions, a thermalized trapped-gas has a truncated Boltzmann distribution and that the trap evaporatively cools as the trapped ensemble thermalizes to repopulate the truncated region and ejects the high energy atoms \cite{KETTERLE1996181}. This results in a semi-equilibrium phase space distribution $f(\textbf{r},\textbf{p})$ that evolves according to the Boltzmann Equation:}
\begin{equation}
 \left(\dfrac{\textbf{p}}{m}\cdot\boldsymbol{\nabla}_\textbf{r} - \boldsymbol{\nabla}_\textbf{r}V \cdot\boldsymbol{\nabla}_\textbf{p} + \pD{}{t}\right)f(\textbf{r},\textbf{p}) = \mathcal{I}_0(\textbf{r},\textbf{p}),
\end{equation}

\noindent{where $V$ is the external potential and $\mathcal{I}_0$ is the collisional integral. The standard treatment is to use the  $s$-wave integral \cite{OJ_Luiten}. However, in the presence of background collisions, we must add the contribution from the background collisions to the collisional integral. This contribution is given by evaluating the time derivative of Eqn. \ref{PDF_Model_Equation} and taking the limit where $t$ goes to zero and using the association:
\begin{equation}
f(\textbf{r},\textbf{p}) = \int_0^U dE \delta\left(E - V(\textbf{r}) - \dfrac{p^2}{2m}\right)f(E).
\end{equation}
\noindent{where $U$ is the trap depth. In this case, the Boltzmann Equation becomes:}
\begin{align}
& \left(\dfrac{\textbf{p}}{m}\cdot\boldsymbol{\nabla}_\textbf{r} - \boldsymbol{\nabla}_\textbf{r}V \cdot\boldsymbol{\nabla}_\textbf{p} + \pD{}{t}\right)f(\textbf{r},\textbf{p}) - \mathcal{I}_0(\textbf{r},\textbf{p}) \label{BackgroundBTE}\\
& =  \Gamma \int_0^U dE\ \delta\left(E - V - \dfrac{p^2}{2m}\right)\left[\int_0^U d\epsilon f(E - \epsilon)P_t(\epsilon) - f(E)\right],\nonumber
\end{align}

\noindent{which we recognize is the statement that changes in the distribution apart from thermalization are due to background collisions. Determining the energy distribution in the thermalization limit is beyond the scope of our work here, but for a known initial phase space distribution $f(\textbf{r},\textbf{p})$, Eqn. \ref{BackgroundBTE} can be integrated numerically and truncated at the maximum trap depth to find an expression for the distribution at later times.}

 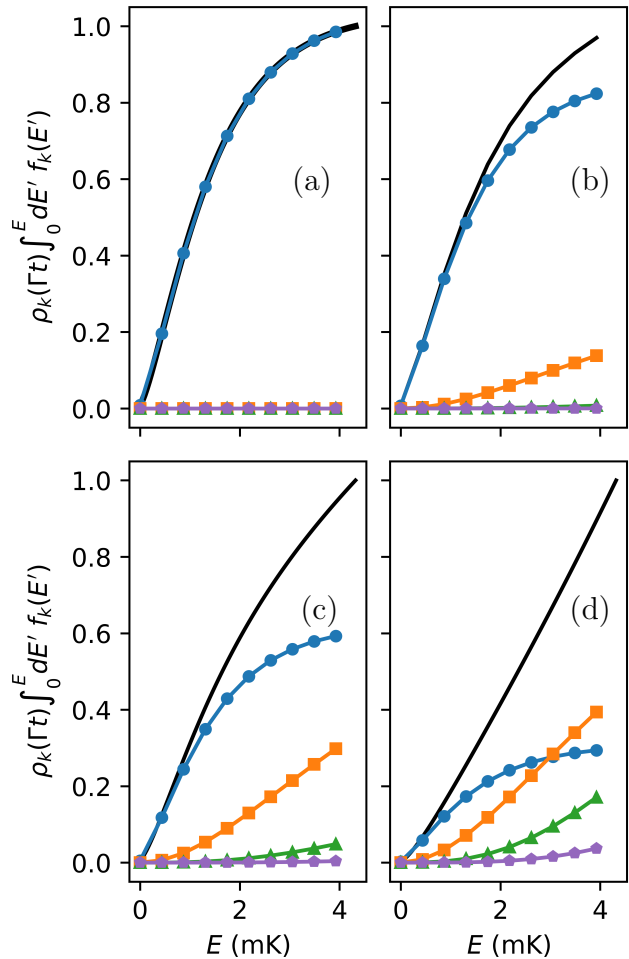
\begin{figure}
	\centering
 \begin{tikzpicture}
    \node at (0,0){\includegraphics[width= \linewidth]{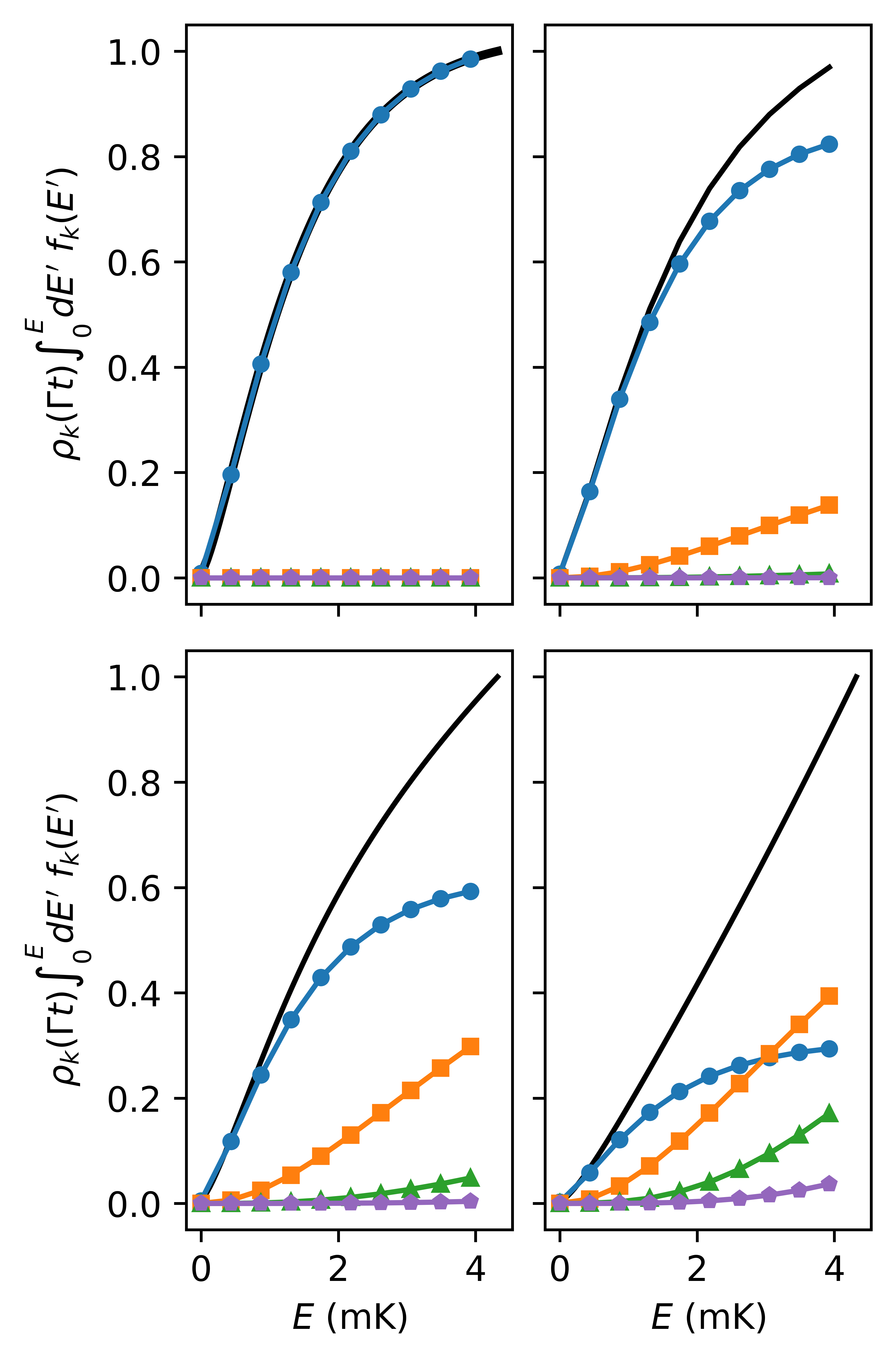}};
    \node at (-0.1,4){\large (a)};
    \node at (3.6,4){\large (b)};
    \node at (0,-1.7){\large (c)};
    \node at (3.6,-1.7){\large (d)};
\end{tikzpicture}

        \caption{Cumulative energy distributions for the collision sub-ensembles. We plot Poisson probability weighted cumulative energy distributions which are the integrals of the $k$-collision sub-ensembles $f_k(E)$ as a function of the energy $E$ calculated using Eqn. \ref{fkEquation} weighted by the Poisson probability $\rho_k$ in Equation \ref{PoissionWeights} at different values of $\Gamma t$. Our initial distribution is a Maxwell-Boltzmann distribution for an $^{87}$Rb gas at $T = 300\ \mathrm{\mu K}$, and we use the transfer energy distribution for the $^{87}\text{Rb}+\text{Ar}$ interaction discussed in Section III. (a) At $\Gamma t = 0$, the total distribution (black) is equal to the initial distribution $f_0(E)$ (blue circles) because the ensemble has not yet undergone collisions. (b) At $\Gamma t = 1$, the ensemble is still dominated by $f_0(E)$, but there is also a small contribution from $f_1(E)$ (orange) and a much smaller contribution from $f_2(E)$ (green). We see the shape of the total distribution as a result of these new contributions. (c) At $\Gamma t = 3$, there are a similar number of particles in the $f_0(E)$ and $f_1(E)$ sub-ensembles. The contribution of the $f_2(E)$ has increased, and $f_3(E)$ (purple) is nonzero. (d) At $\Gamma t = 8$, the distribution is now dominated by the $f_1(E)$ sub-ensemble, and $f_0(E)$ and $f_2(E)$ have approximately equal populations. The total distribution has now changed significantly from the initial distribution, and only reflects $f_0(E)$ in the small energy regime.}
	\label{Sub-Ensemble_Plots}
 
\end{figure}

\section{The Energy Transfer Distribution}

\noindent{We now discuss the energy transfer probability distribution from a thermal background to a trapped particle assuming elastic scattering.  We will neglect inelastic collisions in the transferred energy distribution, because we assume that all inelastic collisions generate trap loss.  We discuss how to calculate $P_t(E)$ using interaction potentials and quantum scattering calculations.  

\begin{figure}
    \centering
    \includegraphics[width=1\linewidth]{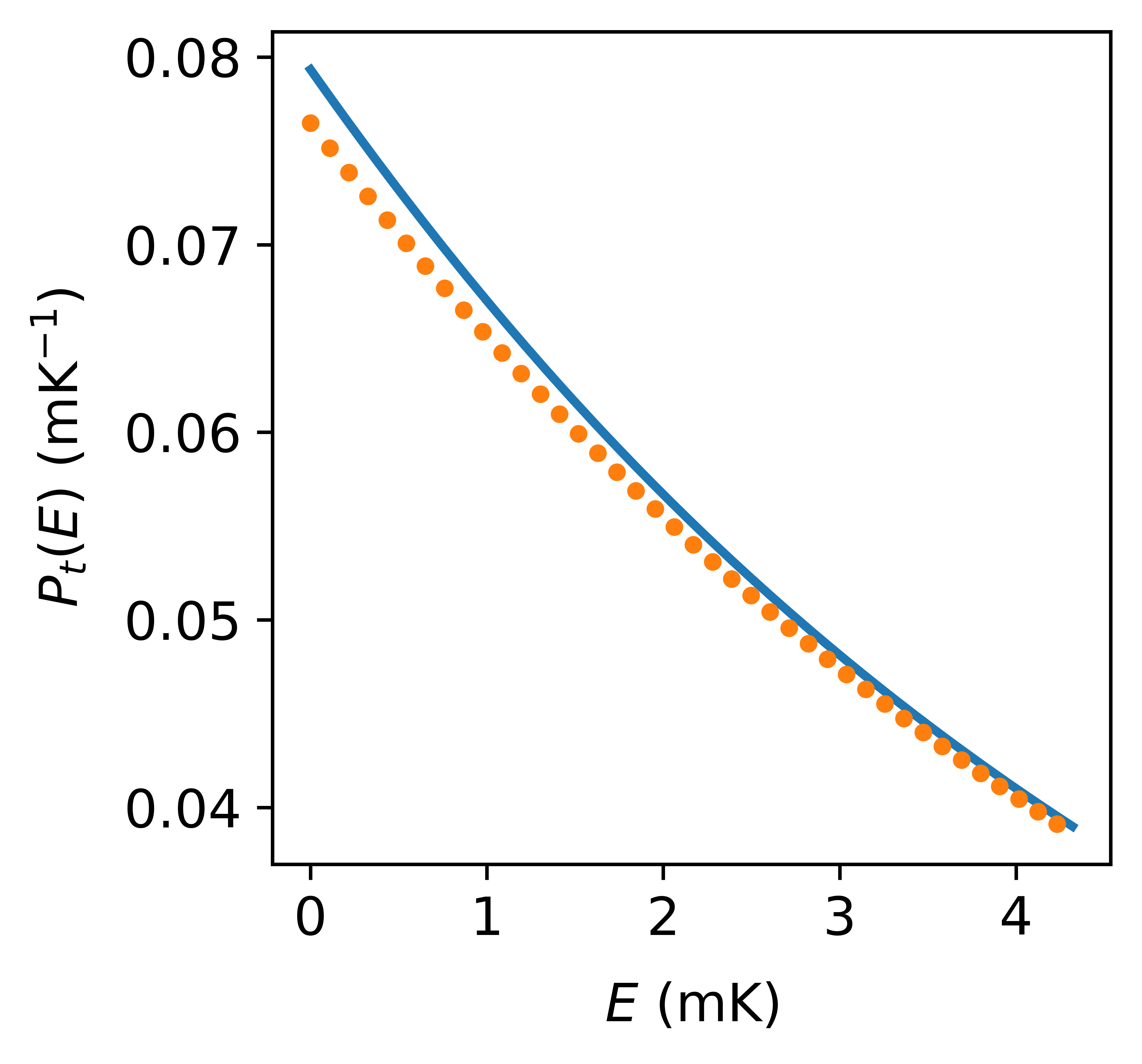}
    \caption{The transfer energy distribution, $P_t(E)$ for rubidium-87 atoms subject to collisions by argon atoms at room temperature.  The solid blue line is $P_t(E)$ obtained from Eqn.~\ref{PtEquation} using the scattering amplitudes computed by quantum scattering calculations as input. The orange dotted line is obtained by differentiating the trap depth dependent loss rate based on quantum diffractive universality (shown in Eqn.~\ref{PtPQDU}) with parameters found in  Ref \cite{Shen_2021}.  These distributions are not normalized over the range of energies shown, reflecting the fact that collisions can cause loss. The two curves are very similar;  the largest deviation between the two is 3.6\%, and the cumulative difference between the them is 2.28\%. Performing analysis in Section V to measure a collision rate $\Gamma$ using the approximate $P_t(E)$ given by quantum diffractive universality instead of the more accurate one from quantum scattering calculations results in an error of 0.61\% in the extracted value of $\Gamma$.  Since the quantum scattering calculations require significant computational resources, it may be more convenient and sufficiently accurate to simply use the universal polynomial.
    }
    \label{PtCurves}
\end{figure}

\subsection{Transfer probability calculations}

\noindent{We consider a trapped particle of mass $m_t$ and velocity $\textbf{v}_t$ subject to an incident background particle of mass $m_b$ and velocity $\textbf{v}_b$. We can reduce the analysis of the collision to the case of single incident particle in the center of mass frame with reduced mass $\mu = m_t m_b/(m_t + m_b)$ and velocity $\textbf{v} = \textbf{v}_t - \textbf{v}_b$ subject to the interaction potential. Consider such a particle in the plane-wave state $\ket{\textbf{k}}$ with an initial momentum $\hbar\textbf{k}$. The probability amplitude of the particle transitioning into a state $\ket{\textbf{k}'}$ with momentum $\hbar\textbf{k}'$ is given by the scattering amplitude $\mathcal{F}(\textbf{k},\textbf{k}')$. Because we are considering elastic collisions in the center of mass frame, this amplitude is only nonzero if $|\mathbf{k}|=|\mathbf{k'}|$.}\\
	
{We assume that the potential energy surface of the interaction potential is spherically symmetric. In this case, the scattering amplitude must also be rotationally invariant, and it can be written as}
	
	\begin{equation}\label{ScatteringAmplitudePLSum}
		\mathcal{F}(\textbf{k},\textbf{k}') = \sum_{\ell = 0}^{\infty} (2\ell + 1) \mathcal{F}_\ell(k) P_\ell(\cos\theta),
	\end{equation}
	
\noindent{where $\theta$ is the angle between $\textbf{k}$ and $\textbf{k}'$, $P_\ell$ are the Legendre polynomials, and the partial wave amplitudes $\mathcal{F}_\ell$ are functions we must compute. We can re-write the scattering amplitudes as functions of the incoming velocity of the reduced mass particle and the scattering angle as}
	\begin{equation}
		\mathcal{F}_{v}(\theta) = \sum_{\ell = 0}^{\infty}  (2\ell + 1) \mathcal{F}_\ell(\mu v/\hbar) P_\ell(\cos\theta).
	\end{equation}
	
\indent{The energy transferred from the background particle to the trapped particle depends on the incident velocity $v$ and the angle $\theta$ by which the reduced mass particle is deflected in the following way \cite{Thomson}}

 	\begin{equation}\label{TransferredE}
		E_t = \dfrac{2\mu^2 v^2}{m_t}\sin[2](\frac{\theta}{2}).
	\end{equation}

 \indent{The probability $\mathcal{P}_v$ that particle with incident velocity $v$ relative to a trapped sensor atom transfers an energy $E$ to that sensor particle is given by the probability that the reduced mass  particle scatters at an angle $\theta$ corresponding to that energy normalized by the total probability that the collision occurs. This normalization constant is the total velocity cross-section $\sigma(v)$. Then, $\mathcal{P}_v$ can be written as}
  
    \begin{equation}\label{Pv_equation}
		\mathcal{P}_v(E) = \dfrac{\abs{\mathcal{F}_v(\theta)}^2}{\sigma(v)} \delta\left(E - \dfrac{2\mu^2 v^2}{m_t}\sin^2\left(\dfrac{\theta}{2}\right)\right).
	\end{equation}

    {To convert this into the ensemble energy transfer distribution, we must average over  the relative velocity distribution. As shown in Appendix D, this  can be very well approximated by the Maxwell Boltzmann distribution that describes the background gas in the rest frame. We also account for the difference in the fluxes in the background gas by adding a velocity weight. Furthermore, the probability of a collision occurring depends on the velocity, so we must multiply by a velocity-dependent cross section that cancels out with the one in the denominator of Eqn. \ref{Pv_equation}. We normalize by the total ensemble averaged velocity-weighted cross-section to ensure unit probability when the distribution is integrated over all energies. This gives us the expression}
	\begin{align}\label{PtEquation}
		P_t(E) = & \dfrac{4\pi}{\langle \sigma(v)v \rangle} \left(\dfrac{m_b}{2\pi k\sn{B}T}\right)^{3/2} 
 \int_0^\infty dv\ \int_0^\pi d\theta\ \sin\theta\ \\ &\times v^3 e^{-m_b v^2/2k\sn{B}T} \abs{\mathcal{F}_v(\theta)}^2 \delta\left(E - \dfrac{2\mu^2 v^2}{m_t}\sin^2\left(\frac{\theta}{2}\right)\right)\nonumber.
	\end{align}

{To find the energy transfer distribution function $P_t(E)$, for our comparison with measurements of a $^{87}$Rb sensor ensemble exposed to an Ar gas, we first calculated the scattering amplitudes $\mathcal{F}_v(\theta)$ using the interaction potential for Rb+Ar from Ref.~\cite{Medvedev2018} as input to a single channel quantum scattering computation as described in Ref. \cite{Booth2019}. We then used these values as input to Eqn.~\ref{PtEquation} to find $P_t(E)$.  The results are shown in Figure \ref{PtCurves}.}\\

{As an aside, we note that these calculations can be very computationally intensive. The calculations can be vastly simplified using quantum diffractive universality, which provides an approximate polynomial expression for $P_t$ that is very simple to calculate \cite{Booth2019}. As shown in Figure \ref{PtCurves}, this polynomial is a good approximation for several species. Details on quantum diffractive universality and the universal polynomial can be found in Appendix E.}

\section{Experimental Measurement}

\subsection{Apparatus and Measurement Procedure}

\noindent{The experimental apparatus used in this study is described in Ref.~\cite{Booth2019}.  Briefly, it consists of a Rb source, a two-dimensional (2D) magneto-optical trap (MOT) section, a three-dimensional (3D) MOT section, and a pump section. The experiment procedure begins by loading Rb atoms into the 2D MOT from the Rb vapor generated by the atom source. A laser beam aligned to the center of the 2D MOT pushes atoms through a differential pumping tube into the 3D MOT region where they accumulate. The magnetic field configuration of the 3D MOT is a spherical quadrupole and is operated with an axial gradient of 13.6 G/cm along the vertical direction. In both the 2D MOT and 3D MOT, the trapping lasers are detuned 12 MHz below the $^{87}$Rb $5^2S_{1/2}\to 6^2P_{3/2}, F=2-3'$ transition, and the `repump' lasers are tuned on resonance with the $5^2S_{1/2}\to 5^2P_{3/2}, F=1-2'$ transition. After the particles are loaded into the 3D MOT, the pump laser frequency is detuned to 60 MHz below the $F =2\to F= 3$ transition for 50 ms to further cool the atom ensemble. Next, the atoms are optically pumped to the $F = 1$ hyperfine ground state by shutting off the repump laser and leaving the pump laser on for another 4 ms. Then, we extinguish the pump laser and ramp up the magnetic field axial gradient to 272 G/cm in 10 ms. Only atoms in the low-field seeking state, $\left.|F = 1, \ m_F = -1\right>$ are confined in the magnetic trap (MT).}\\

{The atoms are held in the MT for a time $t$ before being recaptured in the MOT using the same detunings during the stage just before hyperfine pumping. To count the atom number, we image the MOT ensemble fluorescence onto a photodetector inducing a voltage that is proportional to the atom number $N$. The fraction of atoms $R$ in the MT recaptured in the MOT is}
\begin{equation}
R(t) = \frac{V_{\rm{MT}}(t)}{V_{\rm{MOT}}},
\end{equation}

where $V_{\rm{MT}}(t)$ is the recaptured atomic fluorescence and $V_{\rm{MOT}}$ is the initial MOT fluorescence To set the magnetic trap depth for the atom ensemble, an RF field is applied at the end of the hold time. The RF field induces transitions of the atoms from the trapped state ($\left.|F = 1, \ m_F = -1\right>$) to the $\left.|F = 1, \ m_F = 0,+1\right>$ states that are not trapped by the MT. During the RF radiation phase, the RF frequency is swept between $\nu_{\rm{min}}$ and $\nu_{\rm{max}}$. Atoms with the magnetic energies above $E_m=\hbar\nu_{\rm{min}}$ encounter a resonant RF field and are flipped into an untrappable spin state and escape from the trap.  We therefore define $E_m$ to be the maximum trap depth.  The RF radiation is swept sufficiently for a sufficiently long time in comparison to the motion of the particles in the trap to ensure that every particle which reaches the magnetic energy $E_m$ during their motion will encounter the RF field. As a result, all particles that achieve a maximum magnetic energy of $E_m$ or higher during their evolution are ejected.\\

{It should be noted that the maximum magnetic energy achieved by the particles is not necessarily the total energy of the particle. Through the course of a particle's trajectories, its energy is divided between magnetic potential energy,  gravitational potential energy, and kinetic energy. It is possible that at the time where the particle reaches its maximum magnetic energy, it has nonzero gravitational and kinetic energies. However, in the limit of large magnetic field gradients, these energies will be negligible relative to the magnetic energy. We therefore conduct our experiments at a gradient that is sufficiently high such that the maximum magnetic energy of the particle is essentially the same as the total energy of the particle.  This limit is reached when the trapping gradient is much larger than the minimum gradient required to magnetically levitate the atoms and when the initial kinetic energy of the atoms (set by the MOT) is much smaller than their magnetic potential energy set by the initial cloud size and the MT gradient.  The study of the relationship between the total energy and the maximum magnetic energy outside of this limit is beyond the scope of this work and is the topic of future work.}\\

{To measure the time-dependent cumulative energy distribution function (CDF), $F(E_m,t) = \int_{0}^{E_m} f(E,t) dE$, of the atoms in the magnetic trap, we measure the fraction $R(t)$ of atoms in the magnetic trap that are recaptured in the MOT after the hold time $t$ as a function of the trap depth $E_m$. To observe the time evolution of the cumulative energy distribution, we repeat the CDF measurement for a series of hold times $t\sn{1},...,t_k$. The measured time-dependent CDFs are presented as discrete dots in Fig. \ref{GPR_vs_Experiment}.}

\subsection{Gaussian Process Regression}



\noindent{Statistical and systematic noise present a significant challenge to analyzing our experimental data. Because of this noise, differentiating a raw measured CDF to obtain a probability density function (PDF) leads to negative probabilities and jagged features. Previous analyses guessed an approximate functional form of the CDF and fit the measured data to this form \cite{Shen_2023}. However, because the ensemble does not thermalize, common distributions, such as the Maxwell-Boltzmann distribution, do not accurately describe the system.  It is possible to fit the data to a more realistic distribution, but such methods must incorporate experimental variables (the beam widths, the cloud position, etc) and result in a very complicated fit.}

\indent{Instead of performing a fit, we apply Gaussian Process Regression (GPR) based on the gpytorch Gaussian process library provided by Ref. \cite{gpytorch2018} to calculate a smooth function that describes the CDF. Following Ref \cite{Lin2014}, we impose the physical constraint that the CDF must be monotonically increasing by projecting our calculated function values into the space of monotonically increasing functions. This gives us a continuous, positive-definite PDF. An example of this smoothing procedure applied to experimental data is shown in Figure \ref{GPR_vs_Experiment}. To account for the experimental uncertainty, we use stochastic variational inference -- we re-sample each experimental data point used to calculate the loss in each iteration of the algorithm from a Gaussian distribution centered on the measurement with a standard deviation equal to the uncertainty of the measurement. This combines the fitting uncertainty and experimental uncertainty for the initial distribution that can be used for measurement uncertainty analysis.}\\

\begin{figure}
    \centering
    \includegraphics{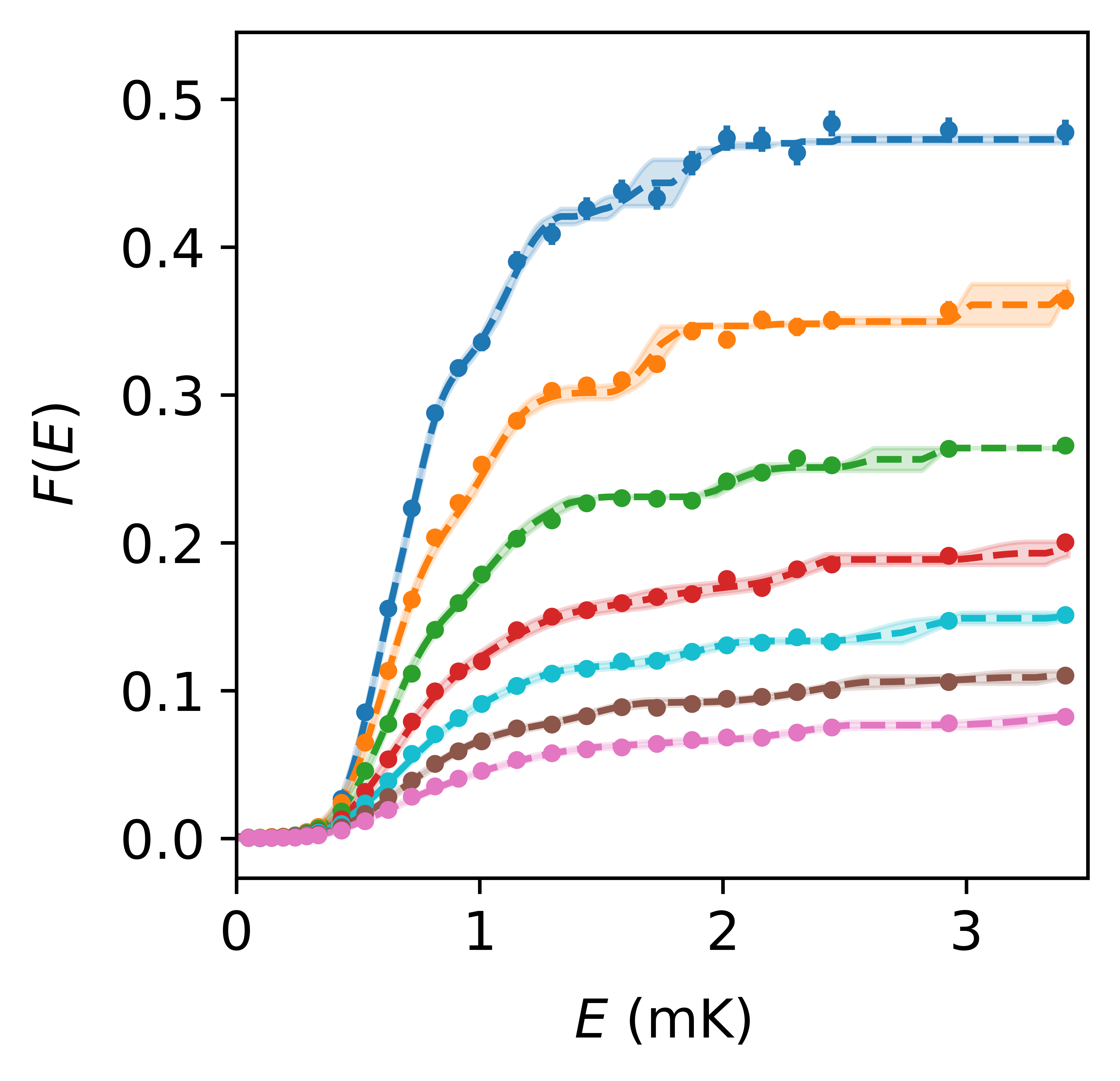}
    \caption{A comparison of Gaussian process regression (GPR) smoothed data to raw experimental data. We have plotted the cumulative distribution function $F(E)$ against the energy $E$ for the experimental data (points) and the corresponding output of the monotonic GPR smoothing procedure (dashed lines). The shaded region denotes the combined uncertainty in the GPR algorithm and experimental measurement. The hold times for each curve were 0.7 s (blue, uppermost), 1.2 s (orange, second highest), 1.8 s (green, third highest), 2.3 s (red, fourth highest), 2.8 s (cyan, fifth highest), 3.4 s (brown, sixth highest), and 3.9 s (pink, seventh highest). We see that the experimental data is jagged and not monotonic, but the GPR output is smooth and strictly increasing. We can then differentiate these curves and obtain a well-defined, physical probability density function. Notice that the GPR output in regions where the data is already smooth and monotonic is very close to the experimental data. Furthermore, regions where the GPR output differs from the data results in a larger uncertainty than regions where the two are close.}
    \label{GPR_vs_Experiment}
\end{figure}

{A major benefit of using GPR to extract a CDF is that the functional nature of the smoothing procedure corrects our experimental measurement values and uncertainties using the context of the other measurements. Because we cannot have negative probabilities, the CDF must increase with the energy. In addition, because our magnetic field and velocity distributions are continuous, we must have a continuous PDF and smooth CDF. Any data points that do not follow this rule must be, either from systematic or statistical effects, far from the mean value that would be measured in the limit of infinite measurements and vanishing systematic error. A smoothing method that uses the experimental data as nodes would miss this detail, and would treat the experimentally measured value as accurate and precise, decreasing the accuracy of the extracted CDF. GPR, on the other hand, uses the experimental data to provide likelihood estimates for the CDF function itself, which means the GPR procedure simultaneously considers the effect of each data point \cite{rasmussen:williams:2006}. As a result, outlying points can be identified and accounted for, mitigating the effect that a single deviation can have on the overall measurement. Furthermore, these outlying points can be identified to have an accordingly larger uncertainty, which reduces the error that would result when fitting to the data. Appendix F contains a more detailed mathematical description of GPR and how it has been applied here.}

\section{Validation of the Energy Distribution Evolution Model}

\noindent{For the comparison of the theoretical model and experiment, we flow pure Ar gas into the experiment chamber through the pump section in the apparatus.  We maintain the Ar pressure at  $10^{-6} \; \rm{Pa}$, which is two orders of magnitude higher than the initial ambient gas pressure. Therefore, we can approximate the entirety of the background to be Ar gas. We measured recapture fractions using the procedure described in the previous section at 28 trap depths and 7 hold times. We performed three repetitions of the experiment.}\\

Since our experimental measurements are of the CDF of the trapped ensemble, we integrate Eqn.~\ref{ModelExpression} to find the time-dependent cumulative energy distribution $F$ of the trapped ensemble, which yields}

\begin{align}\label{Model_CDF}
F(E,t) = \int_0^{E} & dE_0\ \Theta(U - E_0)\sum_{k = 0}^{\infty} \dfrac{(\Gamma t)^k e^{-\Gamma t}}{k!}\\ &\times \prod_{j = 1}^{k} \left(\int_0^U dE_j P_t(E_{j-1} - E_j)\right) f_0(E_k)\nonumber.
\end{align}

\noindent{where $U$ is the maximum possible energy a trapped particle can have. To make the comparison, we must first measure the initial distribution $f_0(E)$ at time $t_0$, then evaluate the expression for $F(E,t)$ and compare it to the experimentally measured recapture fractions $R(E,t+t_0)$ at various hold times.  The only free parameter in our procedure is the total collision rate $\Gamma$, which is determined from the observed loss rate and the ensemble distribution changes. This method provides a much more precise determination of the total collision rate than the method previously used -- an extrapolation of the exponential decay fits back to zero trap depth.}\\

{As an input, our distribution evolution model in Eqn. \ref{ModelExpression} requires a PDF. However, as we have explained in the previous section, converting the experimental data directly to a PDF results in an unphysical distribution. We therefore smooth our shortest hold time ($t_0 = 0.7$ seconds) CDF using GPR, and differentiate the result to obtain an initial distribution $f(t_0)$. We then use Eqn.~\ref{Model_CDF} along with $P_t$ obtained from quantum scattering calculations to numerically calculate the CDF predicted by the model for different times $t = t_e - t_0$, where the $t_e$ are the run times of the experimental data. We evaluated the series in Eqn. \ref{Model_CDF} up to the fourth order heating term, which has a contribution of less than 0.1\% for the time scales considered. Higher order terms are even smaller and were neglected.}\\

{Given $f_0(E)$ and $P_t(E)$, the only free parameter remaining in the model is  $\Gamma$, the total collision rate.  To find it, we perform a global mean square error fit. Namely, we use Eqn.~\ref{Model_CDF} to compute the predicted CDFs and then evaluate the global sum of the squared residuals between all of the CDFs calculated by the model and the experimental CDFs.  To account for the measurement uncertainty in the experimental points, we sample each experimental data comparison point from a Gaussian distribution with mean value equal to the measured data and standard deviation equal to the measurement uncertainty. We then find the value of $\Gamma$ that minimizes the global sum of the squared residuals. This calculation is then repeated for 200 different sample sets providing 200 distinct values of $\Gamma$.  The mean of these values is our reported $\Gamma$, and the standard deviation tells us the propagation of the measurement uncertainty into our value of the collision rate. We find a value of $\Gamma = 0.649(2)\ \text{s}^{-1}$, which is a relative uncertainty of $\Delta\Gamma \approx 3.1\times 10^{-3}\Gamma$. This can be compared to the collision rate found by extrapolating exponential loss rates to zero trap depth, which yields a collision rate 
 of $0.67(1)\ \text{s}^{-1}$. The loss rate extrapolation value is $3.2\%$ higher than the value extracted by our cumulative distribution evolution model, and the relative uncertainty $1.5\times 10^{-2}\Gamma$ in the exponential extrapolation method is five times larger than that of the cumulative distribution evolution model.}\\

{The precision of our analysis can be further improved by using GPR to smooth the experimental data before we fitting them to our model, Eqn. \ref{ModelExpression}. In this case, we repeat our GPR procedure to extract smooth curves for each of the later time experimental measurements. We use the same procedure to obtain $f_0(E)$ and $P_t(E)$ and calculate time-evolved curves using Eqn. \ref{Model_CDF}. We then fit these time-evolved curves to data points obtained from the GPR smoothed data and use the uncertainties in the GPR procedure to calculate the error propagation into our final fit. We extract a collision rate of $\Gamma = 0.646(1)\ \text{s}^{-1}$. Notice that this refined collision rate agrees with the collision rate extracted from the raw experimental data up to the uncertainties in the two measurements, showing that applying GPR to smooth the data increases the precision without significantly changing the extracted value. A plot of the model CDF fits with this value of the collision rate compared to the experimental data is shown in Figure \ref{Model_vs_Experiment}. The smoothed data collision rate is $3.5\%$ smaller than the one obtained from loss rate extrapolation, and the relative uncertainty of the smoothed evolution model method $1.5\times 10^{-3}\Gamma$ is ten times smaller than the exponential extrapolation method. We note that this increase in precision is not due to using the GPR smoothing to increase our sample size to which we can fit. The fitting of the model to the GPR smoothed data uses fewer points than actually gathered in the experiment, demonstrating that the precision is a product of the method and not the sample size.}

\begin{figure}
    \centering
    \includegraphics{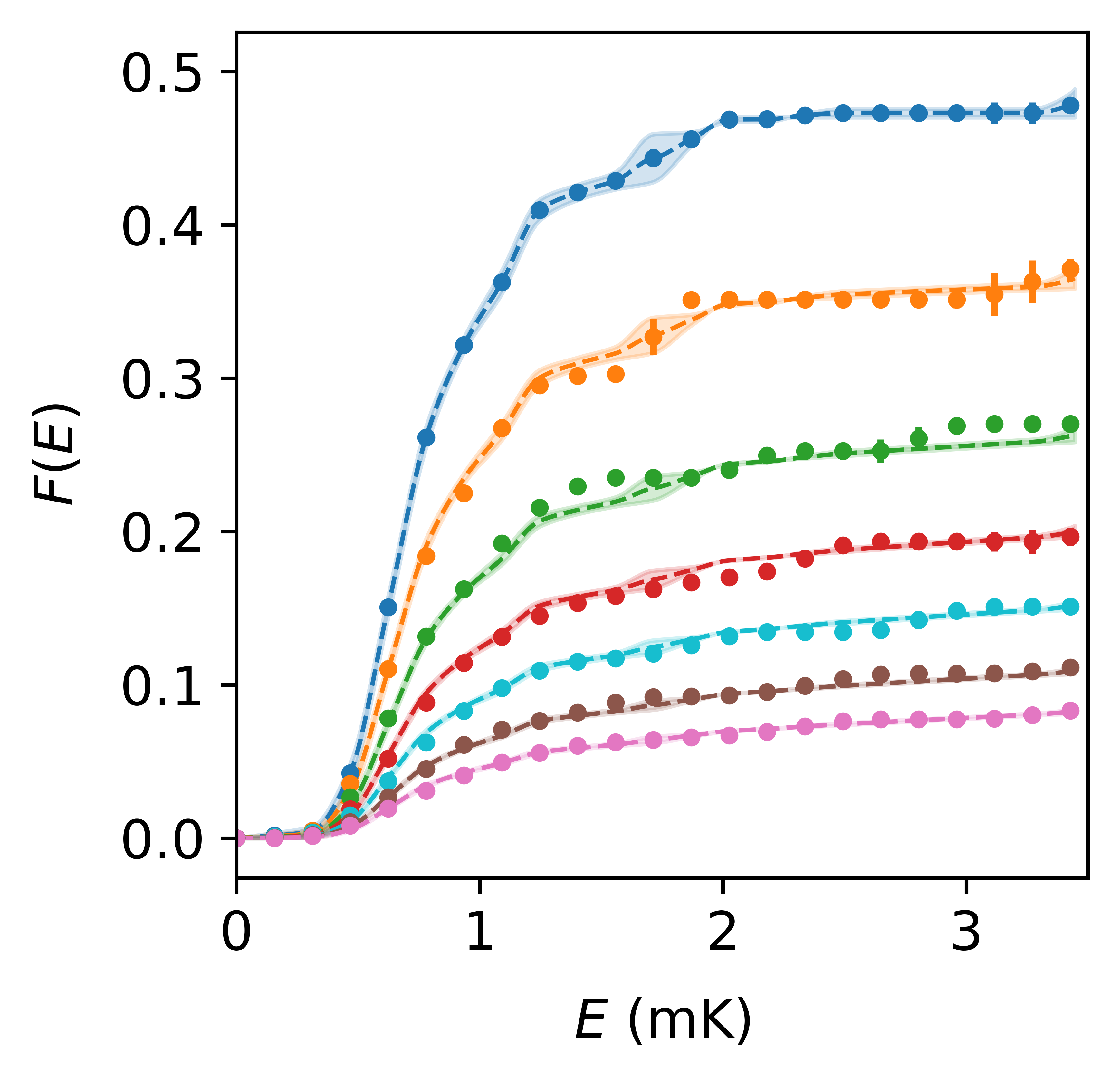}
    \caption{Comparison of the model-predicted cumulative energy distributions at different times to GPR smoothed data. We have plotted the cumulative distribution function $F(E)$ as a function of the energy $E$ for different times. The points are experimental data smoothed by the GPR procedure (these points are the dashed lines in Fig. \ref{GPR_vs_Experiment}), and the dashed lines are the theoretical distributions predicted by the model (Eqn.~\ref{Model_CDF}). The shaded region is the uncertainty in the model that propagates from the uncertainty in the initial distribution. The highest particle number curve (blue, uppermost) was taken at the experimental hold time $t_0 = 0.7\ \text{s}$, and it is used as an initial distribution for the model. The subsequent curves are obtained by evaluating the model at times $t = t_e - t_0$, where $t_e$ is the experimental hold time used for the corresponding CDF. The values of $t_e$ for the curves shown here are 1.2 s (orange, second highest), 1.8 s (green, third highest), 2.3 s (red, fourth highest), 2.8 s (cyan, fifth highest), 3.4 s (brown, sixth highest), and 3.9 s (pink, sevent highest).  The fit of the model to the experimental data provides the collision rate. We find a value of $\Gamma = 0.646(1)\ \text{s}^{-1}$. By construction, we have perfect agreement between the experiment and model at $t = 0$, but we see that the curves agree for longer hold times as well.}\label{Model_vs_Experiment}
 
\end{figure}

\section{Conclusion}

\noindent{We have presented a mathematically rigorous model for the time evolution of the energy distribution of a trapped particle ensemble as a result of collisions with a thermal background gas. We have also described an analysis method that allows us to compare cumulative distributions functions predicted by our model to experimental data. This procedure involves finding the collision-induced transfer energy distribution, using a Gaussian process regression to interpolate the data, and performing a single-parameter fit that extracts the collision rate. We have used this procedure and found good agreement between our model and smoothed experimental data obtained for trapped $^{87}$Rb subject to a thermal argon background. We obtained a total collision rate, using information about heating and trap loss, of $0.646(1)\ \text{s}^{-1}$. This is an uncertainty of below $0.2\%$, which is significantly more precise than any previous measurements. In addition, our model Eqn. \ref{ModelExpression} of the time evolution of the cumulative energy distribution $F(E,t)$ fully corrects measurements for trap heating, a detail that has been neglected or only approximated in previous work.}\\

\indent{This work has several implications for future studies in the area of particle sensors for vacuum metrology. Our work suggests that previous collision rate measurements need to be corrected to account for the change in the distribution, and we have presented a theory to perform this correction that provides the total collision rate with a low uncertainty.  This theory is presently being applied to reanalyze both our previous measurements (Refs.~\cite{Booth2019,Shen_2020,Shen_2021}) and measurements taken by other groups \cite{Eckel2023}. For a thermalized velocity weighted cross-section $\langle\sigma(v) v\rangle$, our collision rate $\Gamma = n\langle\sigma(v)v\rangle$ measurements are equivalent to a measurement of the particle number $n$ and therefore a pressure measurement. As such, our more accurate and precise measurement of the collision rate can be used to make a more accurate and precise pressure measurement.}\\

\indent{We note that this theory can be applied to other species (e.g.~molecules), trap geometries, and experimental procedures, as it does not depend on the distribution of the particles loaded into the trap or the trapping potential that is used. It only depends on one CDF measurement used as an initial distribution $f_0$ to evolve and knowledge of the transfer energy distribution; no additional information about the experimental details is required. All of the dependence of the interaction between the trapped particle and the background comes from the dependence on the transfer distribution, so this method is general to the form of the interaction potential energy surface that is chosen. In addition, while our discussion here focuses on measurements using atomic traps, this analysis holds for any non-interacting cold gas. For example, the same theory extends to a sample that has been sympathetically cooled. As long as a particle number can be measured as a function of time and energy, the details of the cold gas are irrelevant. We further comment that our method is robust against noise in the measurement of $f_0$ and prior heating that has already changed $f_0$ from the distribution first loaded into the trap. We can also generalize the evolution model to systems with multiple background particles by adding an additional set of iterated integrals for each species present.}\\ 

{The analysis techniques we have discussed provide new ways to understand and study diffractive collision physics. In particular, our formulation of the cumulative distribution functions in Eqn. \ref{Model_CDF} suggests a way to measure the transfer distribution using experimental data. This allows us to probe the potential energy surfaces underlying those transfer distributions, and also to assess the accuracy of approximate calculations of the transfer distribution. In Appendix E, we have used our analysis method to assess the approximation of quantum diffractive universality for Rb+Ar collisions, and we intend to extend this analysis to refine our understanding of universality for other species as well.}\\

\section{Appendix}

\subsection{Extrapolating the Loss Rate}

\begin{figure}
    \centering
    \includegraphics[width = \linewidth]{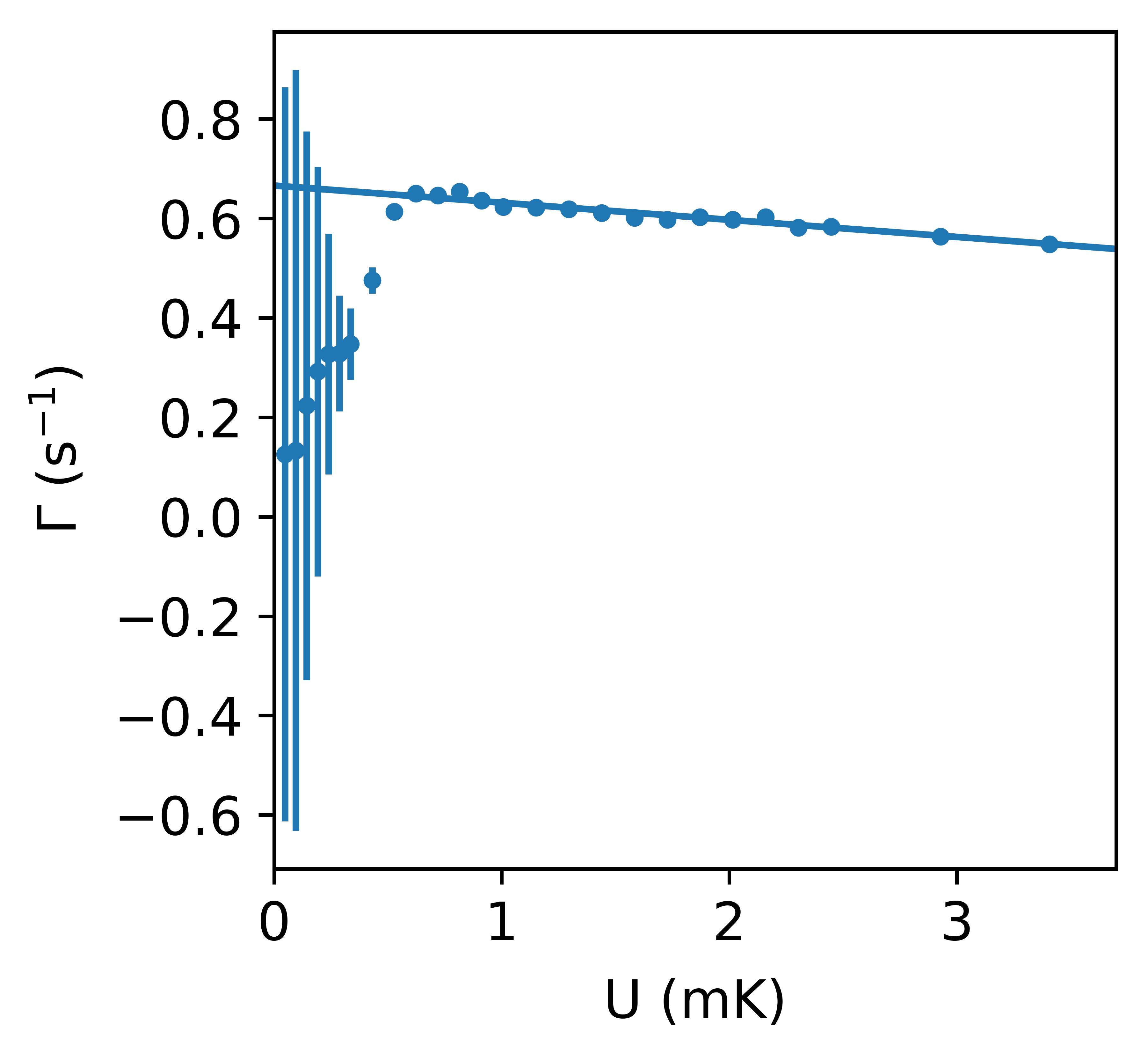}
    \caption{The ensemble loss rate $\Gamma$ versus trap depth $U$ extracted from the data in Fig.~\ref{GPR_vs_Experiment}.  The cumulative particle fraction at a fixed cut energy and different hold times were fit to a simple exponential function to obtain an estimate for the loss rate at different trap depths. These loss rate values were then extrapolated back to find the value at zero trap depth.  The extrapolated value is $\Gamma = 0.67(1) \; \mathrm{s}^{-1}$ if the values at small depths and with large uncertainties are excluded from the fit. If all points are included, the value is 
    $\Gamma = 0.4(3) \; \mathrm{s}^{-1}$. Notice that at low trap depths, the low particle populations result in very large uncertainties.
}
    \label{LossRateVersusTrapDepth}
\end{figure}
\noindent{An alternative approach to the one presented here for obtaining the total collision rate from the observed loss rate is to fit the recaptured fraction as a function of hold time to an exponential decay law at progressively smaller and smaller trap depths.  The results can then be extrapolated to zero trap depth to infer the total collision rate.}\\

To illustrate the limits of this approach, the ensemble loss rate versus trap depth was extracted from the data in Fig.~\ref{GPR_vs_Experiment}.  The cumulative particle fraction at a fixed cut energy and different hold times was fit to a simple exponential function to obtain the loss rate at that trap depth.  The results are shown in Fig~\ref{LossRateVersusTrapDepth}.  The resulting loss rates were then extrapolated back to infer the loss rate at zero trap depth.  The extrapolated value at zero trap depth is $\Gamma = 0.67(1) \; \mathrm{s}^{-1}$ if the values at small trap depth and with large uncertainties are excluded from the fit.  This value is larger by 3.5 \% than the value extracted from the rigorous analysis of the CDFs.\\

There are several problems with this naive approach.  One is that the ensemble loss rate is not a simple exponential and actually increases with time as the ensemble heats.  The second problem is that, because the atoms are not at zero energy, the effective trap depth of the ensemble is not the same as the cut energy at which the CDF is taken.  A manifestation of this is that, because the energy distribution of the atoms has a non-zero width and may also include an offset below which there are no atoms, the trap depth dependence of the loss rate becomes flat at the lowest cut energies.  Evidently, another consequence of this difference between the cut energy and the effective trap depth is that an overestimate results when extrapolating the trap loss rate back to zero cut energy using only data at higher trap depths where the SNR is high.\\

\subsection{Single Particle Heating}

\noindent{In this section, we show our derivation of the energy transferred to a trapped particle after $k$-collisions as shown in Eqn. \ref{K-Collision Energies}. We consider a cold trapped gas of particle mass $m$ in a quadrupole trap subject to the potential $U$ that is only a function of the coordinate $\textbf{x}$. We subject the trapped gas to a background thermal gas of particle mass $m_b$. Consider a trapped particle with momentum $\textbf{p}$ at a position $\textbf{x}$ that collides with a background particle of mass $m_b$, and let $\textbf{p}_t$ be the momentum transferred to the trapped particle by the collision. The final energy $E_f$ of the particle will be}
	\begin{equation}
		E_f = \dfrac{(\textbf{p} + \textbf{p}_t)^2}{2m} + U(\textbf{x}).
	\end{equation}
	\noindent{We now define the transfer energy $E_t$ to be}
	\begin{align}\label{TransferEnergyDefinition}
		E_t = \dfrac{p_t^2}{2m}.
	\end{align}
	
	\noindent{We can rewrite the final energy in terms of the initial energy $E_i$ and the tranfer energy as}
	\begin{equation}\label{One-Collision-Energy}
		E_f = E_i + E_t + \dfrac{\textbf{p} \cdot \textbf{p}_t}{m}.
	\end{equation}
	
	{Next, suppose that the particle with momentum $\textbf{p}_0$ undergoes $n$ collisions with transfer momenta $\textbf{p}_1$,...,$\textbf{p}_n$. Applying induction to Equation \ref{One-Collision-Energy} tells us that the final energy $E_f$ of the particle is given by}
	\begin{equation}
		E_f = E_i + \sum_{k=1}^{n} \left(E_k + \dfrac{\textbf{p}_k}{m}\cdot \sum_{i=0}^{k-1} \textbf{p}_i\right).
	\end{equation}

\subsection{The Thermalization Rate}

\noindent{In our analysis, we do not consider the thermalization of the trapped distribution, that is to say, we assume that the changes in the energy distribution due to internal collisions are negligible. In this section, we shall argue that this assumption is valid by demonstrating that the upper bound on the rate of intratrap collisions (collisions between two trapped particles) is much lower than the background collision rate.}\\

\indent{Previous absorption measurements have shown that our trap has at most of $2.71\times  10^{6}$ atoms. The cloud of trapped sensor particles has a Gaussian profile in each spatial direction with widths $\sigma_x$, $\sigma_y$, and $\sigma_z$, which means the initial spatial distribution $S$ of the cloud is of the form:}
\begin{equation}
S(\textbf{x}) \sim \dfrac{2.71\times 10^{6}}{{(2\pi)^{3/2}\sigma_x \sigma_y \sigma_z }}e^{-x^2/2\sigma_x^2}e^{-y^2/2\sigma_y^2}e^{-z^2/2\sigma_z^2}
\end{equation}
\noindent{where we have normalized the distribution to the total particle number and where we have chosen our coordinates to be centered on the cloud. The peak density of the cloud is the value taken by the distribution at $\textbf{x} = (0,0,0)$, and the density at all other points is strictly lower. The lower bound $\ell_\text{min}$ on the mean free path of each particle is then:}
\begin{equation}
\ell_\text{min} = \dfrac{1}{\sqrt{2}\sigma S(0,0,0)}
\end{equation}

\noindent{where $\sigma$ is the $^{87}\text{Rb} + ^{87}\text{Rb}$ elastic collision cross section. Using the low-velocity elastic cross section from Ref. \cite{Newbury_1995}, and measurements showing that the cloud widths in our setup are approximately $0.7\ \text{mm}$ in the $z$ (axial confinement) direction and $1.4\ \text{mm}$ in the $x$ and $y$ (radial confinement) directions, we find that $\ell_\text{min} \sim 10.4\ \text{m}$. This is almost four orders of magnitude larger than the width of the cloud, indicating that particles are expected to traverse the system thousands of times before undergoing a collision.}\\

\indent{Using our measurement for the initial energy distribution, we can calculate that the average energy $\overline{E} \approx 0.5\ \text{mK}$. Let us assume the limiting case that the energy is entirely kinetic. Then, the maximum velocity $v_\text{max}$ that a particle can have is $\sqrt{(0.5\ \text{mK})k_{\rm{B}}/m}$, where $m$ is the mass of a Rubidium-87 atom. We find that $v_\text{max} \approx 0.22\ \text{m}/\text{s}$. The upper bound on the rate of intratrap collisions is then approximately $v_\text{max}/\ell_\text{min} \approx 0.02\ \text{s}^{-1}$. Using the results from Section V of the main text, the rate of the background collisions is $\Gamma = 0.646(1)\ \text{s}^{-1}$, which is an order of magnitude faster than the intratrap collision rate. In addition, the intratrap collision rate will decrease with time, because particle loss will lead to a lower density. The background collision rate, in contrast, will remain constant, because we maintain a constant background density. Thus, at longer times, the disparity will become even larger. We also comment that the density falls off as a Gaussian away from the center, so a particle that traverses the cloud will experience a lower density and subsequently a slower intratrap collision rate than in our calculation here for most of its trajectory. This verifies our assumption that thermalization can be neglected.}\\

\subsection{The Background Gas Distribution}

\noindent{In this section, we shall justify our approximation of the relative speed distribution of the background particles using the lab frame speed distribution of the background gas. In the lab frame, the background obeys a Maxwell-Boltzmann velocity distribution $d(v)$ with some temperature $T$. However, as the trapped particle is not at rest the relative velocity distribution is some other function $d'(v_r)$. Let $\boldsymbol{\epsilon}$ be the velocity of the trapped particle. The deviation is maximized when $\boldsymbol{\epsilon}$ is collinear with the velocity of incoming background particle, because this maximizes the difference between the magnitudes of the background particle velocity and the relative velocity. We shall then take $\boldsymbol{\epsilon}$ to be in the same direction as the background particle velocity, which reduces this to a one dimensional problem. The relative speed distribution $d'$ is}
\begin{equation}
d'(v) = 4\pi \left(\dfrac{m_b}{2\pi k\sn{B} T}\right)^{3/2} (v+\epsilon)^2 e^{-m_b(v+\epsilon)^2/2k\sn{B} T},
\end{equation}

\noindent{where $m_b$ is the mass of the background particle. Expanding about $\epsilon = 0$, we find}
\begin{align}
d'(v) =  4\pi & \left(\dfrac{m_b}{2\pi k\sn{B} T}\right)^{3/2} v^2 e^{-m_b v^2/2k\sn{B} T} \\ & \times \left(1+\dfrac{2\epsilon}{v} - \dfrac{m_b v\epsilon}{k\sn{B} T} - \dfrac{5m\epsilon^2}{2k\sn{B} T} + \cdots\right)\nonumber,
\end{align}

\noindent{where we have dropped terms of order $\epsilon^k/v^k$, $\epsilon^k/T^k$ and $v^k \epsilon^k/T^k$ for $k \geq 2$. The magnitude of the velocity $\epsilon$ is of order $\sqrt{3k\sn{B} T_t/m_t}$ where $T_t$ is the trap temperature and $m_t$ is the trapped particle mass. Similarly, $v$ is of order $\sqrt{3k\sn{B} T/m_b}$. Then, we can write}
\begin{align}
& \left(1+\dfrac{2\epsilon}{v} - \dfrac{m_b v\epsilon}{k\sn{B} T} - \dfrac{5m_b\epsilon^2}{2k\sn{B} T} + \cdots\right) \\ & \hspace{2cm} \sim \left(1 - \dfrac{15m_b T_t}{m_t T} - \sqrt{\dfrac{m_b T_t}{m_t T}}\right)\nonumber.
\end{align}

\noindent{For our system, $T_t \sim 100\ \mathrm{\mu K}$, $T$ is $294\ \mathrm{K}$, and $m_b/m_t = 0.46$. Then, the correction is of order $1\times 10^{-6}$.}\\

\indent{This analysis is not comprehensive, as the background particle and trapped particle velocities are both drawn from probability distributions that can deviate from the mean value estimates. However, even if we consider $0\ \text{m}/\text{s} < \epsilon < 0.47\ \text{m}/\text{s}$ and $80\ \text{m}/\text{s} < v < 1000\ \text{m}/\text{s}$, which includes 99\% of sampled $\epsilon$ and $v$, the maximum value that the correction term reaches is $0.011$. This also does not account for the fact that the trapped particle velocity is randomly oriented, so the projection of $\boldsymbol{\epsilon}$ along $\textbf{v}$ will be smaller than predicted by the speed distribution.}

\subsection{Quantum Diffractive Universality}\label{QDU Appendix}

\noindent{The quantum scattering calculations required to calculate $P_t(E)$ are often very computationally intensive, as they require a detailed construction of realistic interaction potentials from which the scattering amplitudes are calculated. In the event that the interaction is described by a long-range van-der  Waals type interaction, our calculation can be substantially simplified by making use of quantum diffractive universality (QDU). In Ref. \cite{Booth2019} it was shown that the loss rate $\Gamma_{\rm{loss}}$ of a particle with zero initial energy as a function of the trap depth $U$ normalized to the total collision rate $\Gamma$ can be described by}
 \begin{equation}\label{UniversalPolynomial}
\dfrac{\Gamma_{\rm{loss}}(U)}{\Gamma} = 1 - \sum_{k = 1}^{\infty}\beta_k \left(\dfrac{U}{U_d}\right)^k,
\end{equation}
\noindent{where the $\beta_k$ are species-independent coefficients and $U_d$ is a species-dependent quantum diffraction energy scale. The values of the coefficients for the first six terms and the experimentally measured values of $U_d$ for $^{87}$Rb subject to various background species can be found in Table \ref{BetaTable}  and Table \ref{UdTable} \cite{Shen_2020}.}\\

\phantom{~}\noindent

\begin{table}
    \centering
    \begin{tabular}{p{4cm}p{2cm}}\toprule
        \hspace{0.5cm} Term &\ \ \ \ \  $\beta_i$ \\
        \hline
        \hspace{0.5cm} \ \ \ 1 & \ 0.6730(7) \\
        \hspace{0.5cm} \ \ \ 2 & -0.477(3) \\
        \hspace{0.5cm} \ \ \ 3 & \ 0.228(6) \\
        \hspace{0.5cm} \ \ \ 4 & -0.070(4) \\
        \hspace{0.5cm} \ \ \ 5 & \ 0.012(1) \\
        \hspace{0.5cm} \ \ \ 6  & -0.0009(2) \\ \hline
    \end{tabular}
    \caption{Here, we have listed the first six coefficients $\beta_k$ for the universal polynomial in Equation \ref{UniversalPolynomial} found in Ref \cite{Booth2019}.}
    \label{BetaTable}
\end{table}

\begin{table}
    \centering
    \begin{tabular}{p{4cm}p{2cm}}\toprule
        \hspace{0.5cm} Species & $U_d/k_{\rm{B}}$ (\text{mK}) \\
        \hline
        \hspace{0.6cm} Rb-$\text{N}_2$ &\ \ \ 9.4(1) \\
        \hspace{0.6cm} Rb-He & \ \ \ 32(2) \\
        \hspace{0.6cm} Rb-Ar & \ \ \ 8.8(2) \\
        \hspace{0.6cm} Rb-Xe & \ \ \ 5.00(5) \\
        \hspace{0.6cm} Rb-$\text{H}_2$ & \ \ \ 21.5(6) \\
        \hspace{0.6cm} Rb-$\text{CO}_2$\ \  & \ \ \  8.3(2) \\ \hline
    \end{tabular}
    \caption{This table lists the universal diffractive energies $U_d$ for a trapped Rb ensemble subject to different background species as measured in Ref \cite{Shen_2021}. We have found that the universal approximation works very well for heavier collision partners with a strong van-der Waals interaction character ($\text{N}_2$, Ar, Xe, and $\text{CO}_2$), but is not as effective for the lighter species with a weaker van-der Waals interaction character (He and $\text{H}_2$) \cite{Shen_2023}. Current work is underway to refine these values.}
    \label{UdTable}
\end{table}

{If the trapped ensemble distribution is localized to a single energy, the normalized loss rate is the probability that a collision will transfer an energy greater than $U$. In other words, it is the complement of the cumulative energy transfer distribution. Thus, we can obtain $P_t(E)$ by differentiating Eqn.~\ref{UniversalPolynomial}, which yields}
\begin{equation}\label{PtPQDU}
P_t(E) = \dfrac{1}{U_d}\sum_{k = 0}^{\infty} (k+1)\beta_{k+1} \left(\dfrac{E}{U_d}\right)^k.
\end{equation}

\noindent{This expression is easy to evaluate, and, as shown in Figure \ref{PtCurves}, the $\beta_k$ and $U_d$ values taken from Refs.~\cite{Booth2019,Shen_2020}
provide a good approximation to the transfer distribution found using quantum scattering calculations for Rb+Ar collisions. We repeated the collision rate analysis described in Section V using the approximate $P_t(E)$ polynomial, and obtained a total collision rate of  $0.642(1)\ \text{s}^{-1}$. This is a discrepancy of $0.61\%$ when compared to the collision rate found using the more accurate $P_t(E)$ found using quantum scattering calculations, which could be a tolerable error for uses that do not require precision below the 1\% scale.}\\

\subsection{Gaussian Process Regression}\label{GPR Appendix}

\noindent{While a rigorous mathematical discussion of GPR is far beyond the scope of this paper, we shall briefly describe some introductory principles here and provide Ref. \cite{rasmussen:williams:2006} for a detailed description. Let $\text{GP}(m,\Sigma)$ denote a Gaussian distribution with mean $m$ and covariance $\Sigma$. For the sets of random variables $A = \{a_1,...,a_n\}$ and $B = \{b_1,...,b_m\}$, we define the covariance matrix $\Sigma(A,B)$ to be the $m\times n$ matrix with entries $\Sigma_{ij}$ being the covariances $\Sigma(a_i,b_j)$. A Gaussian process is defined to be a collection of random variables $X$ such that every joint distribution of a finite subset $U\subseteq X$ is a multivariate Gaussian distribution $\text{GP}(m,\Sigma(U,U))$. Subtracting out the mean of the sample makes $m$ vanish, so we can, without loss of generality, consider collections $X$ such that the distribution over a finite $U\subseteq X$ is $\text{GP}(0,\Sigma(U,U))$.}\\

{Let $X = \{x\sn{1},...,x_k\}\in \mathbb{R}^k$, denote some set of input values and let $Y = \{y\sn{1},...,y_k\}\in \mathbb{R}^k$ denote the set of corresponding output values with the corresponding set of uncertainties uncertainties $\epsilon = \{\epsilon\sn{1},...,\epsilon_k\}\in \mathbb{R}^k$. Consider the Gaussian distribution $\text{GP}(Y,\Sigma(X,X)+\epsilon \epsilon^t)$, where $\Sigma:\mathbb{R}\times \mathbb{R}\to \mathbb{R}$ is some function such that $\Sigma(X,X)$ is a valid (symmetric and positive definite) covariance matrix and $\epsilon \epsilon^t$ is the outer product matrix with entries $(\epsilon \epsilon^t)_{ij} = \epsilon_i \epsilon_j$. We can now sample an element $f(X)\in \mathbb{R}^k$ with entries $f_i$ from this distribution. Notice that by associating $f(x_i) = f_i$, we have defined a function $f:X\to\mathbb{R}$. We can now choose a new set $X^\ast = \{x\sn{1}^\ast,...,x_m^\ast\}\in \mathbb{R}^k$ of test inputs. Our goal is  to extend our existing function $f$ to another function over $X\cup X^\ast$. To do this, we want to find the probability distribution of values of $f(x_i^\ast)$ that agree with the initial points $f(X)$, or the conditional probability $p(f(x_i^\ast)|X,X^\ast,f(X))$. As a prior probability, we take the one generated by the function $\Sigma$, which means the joint distribution of $f(X)$ and $f(X^\ast)$ is:}

\begin{equation}
\twovector{f(X)}{f(X^\ast)} \sim \text{GP}\left(\twovector{Y}{0},\fourmatrix{\Sigma(X,X)}{\Sigma(X,X^\ast)}{\Sigma(X,X^\ast)}{\Sigma(X^\ast,X^\ast)}\right).
\end{equation}

\noindent{Using the properties of Gaussian distributions, we can calculate that the conditional probability distribution $p(f(X^\ast)|X,X^\ast,f(X))$ is the Gaussian process $\text{GP}\big(\Sigma(X^\ast,X)\Sigma(X,X)^{-1}f(X),\Sigma(X^\ast,X^\ast)-\Sigma(X^\ast,X)K(X,X)^{-1})\Sigma(X,X^\ast)\big)$. Notice that $X^\ast$ along with the mean and standard deviation for this distribution provide additional input data, output data, and uncertainties over which we can iterate. That is, we can take $X_1 = \{X,X^\ast\}$ as our input data, $Y = \{Y,\Sigma(X^\ast,X)\Sigma(X,X)^{-1}f(X)\}$ as our output data, and $\epsilon_1 = \{\epsilon,\Sigma(X^\ast,X)K(X,X)^{-1})\Sigma(X,X^\ast)\}$ as our uncertainty.}\\

{Choosing test points and a function $\Sigma$, commonly called the kernel, is of crucial importance to GPR, but it is a subject far beyond the scope of our discussion. We instead point the reader to Ref. \cite{gpytorch2018}, and make the comment that we choose kernels such that $\lim_{x\to x'}\Sigma(x,x') = 1$ and $\lim_{x-x'\to\infty} \Sigma(x,x') = 0$. This imposes the condition that points that are close together are highly correlated and points that are far apart are not correlated, which means our function is smooth. Following Ref \cite{Lin2014}, we impose the monotonicity constraint by projecting the calculated mean functions and their uncertainties into the space of monotonically increasing functions. This is accomplished by defining a metric on the function space and replacing each calculated function with the closest monotonically increasing function. We choose the $L^2$ metric $d$ given by:}

\begin{equation}
d(f,g) = \int_U dx\abs{f(x) - g(x)}^2,
\end{equation}

\noindent{where $U\subseteq \mathbb{R}$ is some set over which we consider the functions $f$ and $g$. We implement this projection by repeatedly sampling functions from the distribution over function space generated by the GPR algorithm, and throwing out all functions that are not monotonoically increasing. While this does not actually sample the entire space of monotonic functions, it is an effective way to sample the subset of monotonic function space that is close to the output of the GPR procedure under the $L^2$ metric.}\\

{Notice that if an experimental data point is close in value to other nearby points and the curve is monotonically increasing there, the mean of the final calculated distribution over functions will be close to that data point. However, if the data point jumps away from the nearby points or is not greater than the previous points due to experimental noise, our procedure will shift the mean distribution point away from it. In this way, we are able to correct for statistical and systematic effects in the data using the context of the other data points.}\\

\section{Acknowledgements}

\noindent{We acknowledge financial support from the Natural Sciences and Engineering Research Council of Canada (NSERC/CRSNG) and the Canadian Foundation for Innovation (CFI). This work was done at the Center for Research on Ultra-Cold Systems (CRUCS). P.S. acknowledges support from the German Research Foundation (DFG) within the GRK 2079/1.}

\bibliography{heating_new}

\end{document}